\documentclass[sigconf, screen]{acmart}
\usepackage[utf8]{inputenc}

\copyrightyear{2024}
\acmYear{2024}
\setcopyright{rightsretained}
\acmConference[KDD '24]{Proceedings of the 30th ACM SIGKDD Conference on Knowledge Discovery and Data Mining}{August 25--29, 2024}{Barcelona, Spain}
\acmBooktitle{Proceedings of the 30th ACM SIGKDD Conference on Knowledge Discovery and Data Mining (KDD '24), August 25--29, 2024, Barcelona, Spain}\acmDOI{10.1145/3637528.3671883}
\acmISBN{979-8-4007-0490-1/24/08}

\title{Reliable Confidence Intervals for Information~Retrieval~Evaluation~Using~Generative~A.I.}

\author{Harrie Oosterhuis}
\affiliation{
    \institution{Google Research}
    \country{}
}
\affiliation{
    \institution{Radboud University}
    \city{Amsterdam}
    \country{The Netherlands}
}
\email{harrie.oosterhuis@ru.nl}
\authornote{Authors contributed equally to this work.}
\authornote{Now at Google DeepMind.}
\authornote{Work done while Harrie Oosterhuis was working at Google Research.}

\author{Rolf Jagerman}
\affiliation{
    \institution{Google Research}
    \city{Amsterdam}
    \country{The Netherlands}
}
\email{jagerman@google.com}
\authornotemark[1]
\authornotemark[2]

\author{Zhen Qin}
\affiliation{
    \institution{Google Research}
    \city{New York}
    \country{US}
}
\email{zhenqin@google.com}
\authornotemark[2]

\author{Xuanhui Wang}
\affiliation{
    \institution{Google Research}
    \city{Mountain View}
    \country{US}
}
\email{xuanhui@google.com}
\authornotemark[2]

\author{Michael Bendersky}
\affiliation{
    \institution{Google Research}
    \city{Mountain View}
    \country{US}
}
\email{bemike@google.com}
\authornotemark[2]

\date{May 2024}

\usepackage{graphicx}
\usepackage{acronym}
\usepackage{xcolor}
\usepackage{bbm}
\usepackage{mleftright}
\usepackage{subcaption}
\usepackage{listings}
\usepackage{pgfplots}
\usepackage[inline]{enumitem}
\usepackage{numprint}

\usepgfplotslibrary{fillbetween}
\usetikzlibrary{calc}

\pgfplotsset{compat=1.16}
\definecolor{gRed}{HTML}{c5221f}
\definecolor{gYellow}{HTML}{f29900}
\definecolor{gGreen}{HTML}{188038}
\definecolor{gBlue}{HTML}{1967d2}
\definecolor{gGrey}{HTML}{9aa0a6}

\pgfplotsset{
    every axis/.append style={
        label style={font=\small},
        tick label style={font=\small},
    }
}

\acrodef{LLM}{large language model}
\acrodef{IR}{information retrieval}
\acrodef{CI}{confidence interval}
\acrodef{DCG}{discounted cumulative gain}
\acrodef{AI}{artificial intelligence}
\acrodef{PPI}{prediction-powered inference}
\acrodef{CRC}{conformal risk control}



\begin{abstract}
    The traditional evaluation of \ac{IR} systems is generally very costly as it requires manual relevance annotation from human experts.
    Recent advancements in generative artificial intelligence --specifically \acp{LLM}-- can generate relevance annotations at an enormous scale with relatively small computational costs.
    Potentially, this could alleviate the costs traditionally associated with \ac{IR} evaluation and make it applicable to numerous low-resource applications.
    However, generated relevance annotations are not immune to (systematic) errors, and as a result, directly using them for evaluation produces unreliable results.
    
    In this work, we propose two methods based on prediction-powered inference and conformal risk control that utilize computer-generated relevance annotations to place reliable \acp{CI} around \ac{IR} evaluation metrics.
    Our proposed methods require a small number of reliable annotations from which the methods can statistically analyze the errors in the generated annotations.
    Using this information, we can place \acp{CI} around evaluation metrics with strong theoretical guarantees.
    Unlike existing approaches, our conformal risk control method is specifically designed for ranking metrics and can vary its \acp{CI} per query and document.
    Our experimental results show that our \acp{CI} accurately capture both the variance and bias in evaluation based on \ac{LLM} annotations, better than the typical empirical bootstrapping estimates.
    We hope our contributions bring reliable evaluation to the many \ac{IR} applications where this was traditionally infeasible.
\end{abstract}

\begin{document}
\maketitle

\acresetall

\section{Introduction}

The evaluation of \ac{IR} systems is an important and long-established part of the \ac{IR} field~\citep{webber2010measurement, harman2011information, voorhees2005trec}.
The goal of standard \ac{IR} systems is to retrieve and rank documents according to their relevance to a query and user.
Accordingly, standard \ac{IR} evaluation metrics (e.g., precision, recall, \ac{DCG}, etc.) measure how relevant the top ranked items are for a set of known queries~\citep{jarvelin2002cumulated, jarvelin2017ir, buckland1994relationship}.
Accordingly, traditional evaluation requires a dataset with examples of documents, queries and annotations that indicate the relevance of documents to queries~\citep{sanderson2010test, voorhees2019evolution, kekalainen2002using, lesk1968relevance}.
Whilst documents and queries are often gathered by logging user interactions, relevance annotations are traditionally created through the labour of human experts, who are trained for the specific labelling task~\citep{nguyen2016ms, craswell2021trec, harman2011information, bailey2008relevance}.
Consequently, creating a new dataset for \ac{IR} evaluation purposes is generally very costly, and as a result, no large datasets have been created for many \ac{IR} settings~\citep{yarmohammadi2019robust, huang2023improving, cui2015multilingual, thomas2023large}.
Thus, for these low-resource settings traditional evaluation is not available in practice.

Despite the large costs involved, there has been a continuous effort, often driven by initiatives like TREC and CLEF, to create new datasets for different \ac{IR} tasks~\citep{van1975document, bonifacio2022inpars, harman2005trec, qin2013introducing, kando1999overview, chapelle2011yahoo, lucchese2018selective, thakur2021beir, voorhees2003overview, qin2010letor, voorhees2021trec, tsatsaronis2015overview, peters2001cross}.
Since the foundational Cranfield collection~\citep{van1975document}, many datasets have been created for ad-hoc retrieval~\citep{harman2005trec, kando1999overview, voorhees2003overview, peters2001cross}.
However, to match the large variety of \ac{IR}-related tasks, many other datasets were subsequently introduced, accordingly;
For example, datasets with numerical \ac{IR} features for learning-to-rank~\citep{chapelle2011yahoo, qin2013introducing, qin2010letor}, or large collections of natural language question-answering examples such as MS MARCO~\citep{nguyen2016ms} and BioASQ~\citep{tsatsaronis2015overview}.
Similarly, recent years have seen the introduction of the TREC-DL~\citep{craswell2021trec}, BEIR~\citep{thakur2021beir} and Istella22~\citep{dato2022istella22}, among others~\citep{bonifacio2022inpars, voorhees2021trec}, specifically for the evaluation of neural \ac{IR} systems.
Consequently, most of the subsequent advancements in neural \ac{IR} were only possible because of the availability of these datasets and the reliable benchmarking that they enable~\citep{mitra2018introduction, onal2018neural, guo2020deep, mitra2017neural}.
This highlights the importance and impact of reliable evaluation on the \ac{IR} field~\citep{harman2011information, sanderson2010test, voorhees2019evolution}.

It is thus no surprise that potential novel data sources are received with great excitement.
In the past two years, new advancements in generative \acl{AI}~\citep{jovanovic2022generative, pavlik2023collaborating, liu2023generative, creswell2018generative, goodfellow2020generative}, especially the arrival of \acp{LLM}~\citep{brown2020language, ouyang2022training, team2023gemini}, are speculated to bring potentially groundbreaking sources for \ac{IR} evaluation~\citep{clarke20234, faggioli2023perspectives}.
\acp{LLM} are trained on extremely large corpora of diverse texts for the task of generating fluent natural language~\citep{tay2022unifying,longpre2023flan}.
Importantly, \acp{LLM} are also capable at performing numerous miscellaneous tasks such as question-answering, text-summarization and text-annotation~\citep{tornberg2023chatgpt,alizadeh2023open,gilardi2023chatgpt, clarke20234, team2023gemini}.
Compared to annotation by human experts, annotation via \acp{LLM} can be performed relatively cheaply and at much larger scale~\citep{thomas2023large, macavaney2023one}.
Several existing studies have already investigated the application of \ac{LLM}-generated relevance annotations to \ac{IR} evaluation~\citep{faggioli2023perspectives, clarke20234, macavaney2023one}.
In particular, \citet{thomas2023large} found that, when applied correctly to specific settings, \acp{LLM} can produce better labels than third-party assessors at a fraction of the costs.
Thus, there is a clear potential for \ac{IR} evaluation based on computer-generated relevance annotations.

However, a fundamental issue with evaluation based on \acp{LLM}, or other generative models, is that they are bound to make errors~\citep{clarke20234, thomas2023large, bender2021dangers}.
Part of these errors are coincidental, since perfect relevance prediction is infeasible in practice, but other errors are systematic~\citep{faggioli2023perspectives}.
For instance, an \ac{LLM} could systematically misestimate relevance in certain domains or on documents with particular attributes~\citep{thomas2023large, bender2021dangers}.
In turn, these errors could affect the final evaluation metrics and result in incorrect assessments of performance. 
Unfortunately, generative models cannot give trustworthy insight into their own reliability~\citep{jovanovic2022generative, pavlik2023collaborating}.
Thus when solely relying on \ac{LLM}-based evaluation, one cannot be certain how reliable their conclusions are.

In this work, we investigate how computer-generated relevance annotations can be used for reliable evaluation, by constructing \acp{CI} around ranking metrics with them~\citep{webber2010measurement, sakai2014statistical}.
Our approach requires a small number of reliable \emph{ground truth} annotations, in order to statistically analyze the distribution of errors that exist in the generated annotations.
Subsequently, we apply two state-of-the-art methodologies~\citep{angelopoulos2022conformal, angelopoulos2023prediction} with a strong theoretical grounding to find reliable \acp{CI}.
In this work, we provide two main methodological contributions:

Our first contribution is the novel application of \acfi{PPI} to \ac{IR} evaluation~\citep{angelopoulos2023prediction}.
\ac{PPI} applies classical methods for building \acp{CI} but builds them around the error between the predicted and true values of a metric.
Thereby, somewhat-reasonable predictions can lead to substantially smaller \acp{CI} than classical \ac{CI} around just the metric value.
The limitations of \ac{PPI} is that it does not utilize the uncertainty of the generative model, and that it only provides a \ac{CI} around the final metric value.

Our second contribution addresses these limitations by proposing a novel \acfi{CRC} approach~\citep{angelopoulos2021gentle, angelopoulos2022conformal}.
We introduce a novel method to place an optimistic and a pessimistic prediction around each generated relevance label, which follows the confidence of the generative model.
These predictions can be propagated to form an interval around metrics on the query or dataset level.
Through \ac{CRC}, our approach calibrates the intervals to guarantee that the true value lies between them with a minimum probability.
In other words, our method puts lower and upper bounds around the relevance of each document that naturally translate to reliable \ac{CI} on query and dataset-level metrics.
Thereby, unlike \ac{PPI}, our \ac{CRC} approach does utilize the confidence of the generative model and can provide \ac{CI} on query-level performance.

Our results on several \ac{IR} benchmarks show that both our methods provide \acp{CI} around \ac{LLM}-based metric predictions that accurately capture the true values, while also being significantly less wide than those of previous \ac{CI} methods~\citep{webber2010measurement, cormack2006statistical}.
Moreover, unlike other approaches~\citep{angelopoulos2023prediction}, our \ac{CRC} method can vary \acp{CI} per document, query or collection of queries and can thus better indicate where a generative model is more or less reliable.

To the best of our knowledge, our novel approaches are the first that leverages computer-generated relevance annotations to produce reliable \acp{CI} for \ac{IR} evaluation.
We hope this contribution opens up novel possibilities for reliable benchmarking of low-resource \ac{IR} tasks that have been traditionally infeasible.
\section{Related Work}
\label{sec:relatedwork}

\subsection{Confidence intervals for IR evaluation}
Evaluation is a well-established core part of the \ac{IR} field~\citep{van1975document,webber2010measurement, harman2011information, voorhees2005trec, peters2001cross}.
Generally, it aims to measure how well a retrieval system can produce a list of ranked documents in response to a user query~\cite{harman2011information, voorhees2019evolution, sanderson2005information}.
The most prevalent form of \ac{IR} evaluation relies on datasets containing example queries, documents and human-annotated relevance labels~\citep{van1975document, harman2011information, voorhees2019evolution, sanderson2010test}.
Accordingly, there is a long history of efforts to create such datasets in the \ac{IR} community, such as TREC~\cite{voorhees2003overview, voorhees2005trec,craswell2021trec,voorhees2021trec, harman2005trec}, CLEF~\cite{peters2001cross}, NTCIR~\citep{kando1999overview} and many others~\citep{van1975document, bonifacio2022inpars, qin2013introducing, chapelle2011yahoo, lucchese2018selective, thakur2021beir, qin2010letor, tsatsaronis2015overview, dato2022istella22}.
Despite the enormous importance of these datasets, they are known to have limitations.
For instance, expert annotators can give conflicting relevance assessments, and the actual users of an \ac{IR} application can  disagree with the experts as well~\citep{sanderson2010test}.
Furthermore, the constructions of these datasets is often costly which puts constraints on their size~\citep{chapelle2011yahoo, voorhees2003overview, van1975document, carterette2006minimal}.
As a result, \ac{IR} datasets can only represent a limited slice of the queries that a real \ac{IR} system receives~\citep{webber2010measurement, carterette2006minimal}.

Accordingly, statistical approaches to \ac{IR} evaluation have been developed to deal with these limitations.
For example, it has become common practice to use significance tests to ensure that observed differences in \ac{IR} metrics are, with high probability, not the result of random chance~\cite{smucker2007comparison, urbano2019statistical, fuhr2018some}.
Confidence intervals (CIs) have been used to express the uncertainty that comes from using the dataset sample of queries to estimate performance over all queries~\cite{cormack2006statistical, sakai2014statistical}.
Furthermore, previous work has also applied \ac{CI} for relevance annotator disagreement~\citep{demeester2016predicting, hripcsak2005agreement} and missing relevance annotations~\citep{webber2013approximate, aslam2006statistical, yilmaz2008simple}.
The statistical methods used to construct \ac{CI} by previous work in \ac{IR} have been based on empirical bootstrapping techniques~\citep{diciccio1996bootstrap, diciccio1988review, hesterberg2011bootstrap}.
To the best of our knowledge, our work is the first to consider \ac{PPI} and \ac{CRC} methods for \ac{IR} evaluation~\citep{angelopoulos2023prediction,angelopoulos2022conformal}.

\subsection{LLMs for relevance annotation generation}
Recent advances in \acp{LLM} have demonstrated impressive capabilities on a broad range of tasks~\citep{tornberg2023chatgpt,alizadeh2023open,gilardi2023chatgpt,team2023gemini}.
Previous work has specifically considered using \acp{LLM} for relevance annotation in an \ac{IR} context~\citep{faggioli2023perspectives, clarke20234, macavaney2023one, thomas2023large}.
\citet{thomas2023large} propose using ground truth relevance labels from human annotators, to find a prompt that results in the most accurate \ac{LLM} generated labels.
They claim that this method produces relevance annotations at the same quality as third-party human assessors but at a fraction of the costs~\citep{thomas2023large}.
\citet{clarke20234} propose that \ac{LLM} relevance-annotation should be approached as a spectrum, since the involvement of humans can be varied.
For instance, one could delegate most work to an \ac{LLM} but add some human verification, as a compromise of reduced costs and reliability.
\citet{faggioli2023perspectives} support this approach, as they see severe risk in blindly following \ac{LLM} generated relevance labels (at least for the current state-of-the-art \acp{LLM}).
The danger foreseen by both is that generated labels can make systematic errors that lead to incorrect and unreliable evaluation of \ac{IR} systems~\citep{bender2021dangers, clarke20234, faggioli2023perspectives}.
Our work addresses this problem, and is thus very related; specifically, our contribution can be seen as an approach of human verification designed to quantify uncertainty stemming from \ac{LLM} usage.

\section{Preliminaries}

\subsection{Evaluation metrics for retrieval systems}

The general approach to the evaluation of a retrieval system is to consider the expected value of a ranking metric across the queries it will receive~\citep{harman2011information}.
Standard ranking metrics assume that each document has certain relevance to a query~\citep{kekalainen2002using}.
For a set of labels $\mathcal{R}$, we use $P(R = r \mid d, q)$ to denote the probability that a human rater would give rating $r \in \mathcal{R}$, to the combination of document $d$ and query $q$.
We define \emph{relevance} as the expected rating value over this distribution:
$
    \mu(d \mid q) = \sum_{r \in \mathcal{R}}P(R = r \mid d, q)\, r.
$
In standard ranking settings, the goal is to place more relevant documents at higher ranks~\citep{jarvelin2017ir}.
Ranking metrics capture this goal by giving a weight to each rank, which indicates how much the relevance of a document placed at that rank should contribute to the metric~\citep{jarvelin2002cumulated}.
We will use $\omega$ to denote our weighting function which takes the rank of a document as its input.
For example, \emph{Precision@K} has the following corresponding weight function:
$\omega_\text{Prec@K}(x) = \frac{1}{K}\mathbbm{1}[x \leq K]$;
and the popular \ac{DCG}~\citep{jarvelin2002cumulated}:
$\omega_\text{DCG@K}(x) = \frac{\mathbbm{1}[x \leq K]}{\log_2(x + 1)}$.
Given a choice of weights and $\mathcal{D}_q$, the set of available documents for query $q$, the metric value for a single query is:
\begin{equation}
    U(q) = \sum_{d \in \mathcal{D}_q}\omega(\text{rank}(d \mid q,\mathcal{D}_q))\,\mu(d \mid q).
    \label{eq:utility}
\end{equation}
Let $P(q)$ denote the natural query distribution;
the performance of a system in terms of the metric is:
\begin{equation}
    U(\mathcal{Q}) = \mathbb{E}_{q\sim\mathcal{Q}}\mleft[U(q)\mright]= \sum_{q \in \mathcal{Q}} P(q \mid \mathcal{Q}) U(q).
\end{equation}
In practice, $U$ can never be computed exactly, since $P(q)$ and $P(R = r \mid d, q)$ are never directly available.
Thus, generally, an estimate of $U$ is made on a large set of sampled user queries and a few relevance judgements per document-query pair~\citep{webber2010measurement, sanderson2010test}.

\subsection{Problem setting}
\label{sec:problemsetting}

In our setting, we make the standard assumption that a large set of sampled user queries and a document collection are available.
However, we do not assume that there are human relevance annotations for every document-query pair, and instead, we assume that ground truth annotations are only available for a small subset: the first $n$ queries out of a total of $N$ queries.
Unique to our problem setting is that a generative model is available to predict relevance annotations.
Furthermore, our aim is not to give a point estimate of the true performance of a system,
instead our goal is to construct a reliable \ac{CI} around the true value of an \ac{IR} metric.
Thereby, we utilize the generated relevance annotations, but still explicitly indicate the resulting uncertainty in our evaluation with \acp{CI}.

In formal terms, let $\alpha \in [0,1]$ be a confidence parameter, we desire to find a lower bound $\hat{U}_\text{low}$ and an upper bound $\hat{U}_\text{high}$ s.t:
\begin{equation}
    P\big(\hat{U}_\text{low} \leq U(\mathcal{Q}) \leq \hat{U}_\text{high} \big) \geq 1 - \alpha.
\end{equation}
Accordingly, $\alpha$ can be chosen to match the desired confidence, i.e., $\alpha = 0.05$ leads to a $95\%$ \ac{CI}.
Additionally, in Section~\ref{sec:method:conformal}, we propose a \ac{CRC} method that can also bound the performance per query, thereby, it can meet the following query-level \ac{CI} goal:
\begin{equation}
    P\mleft(\hat{U}_\text{low}(q) \leq U(q) \leq \hat{U}_\text{high}(q)
    \mid q \sim \mathcal{Q} \mright)
    \geq 1 - \alpha.
\end{equation}
We assume that the available generative model predicts a distribution over possible relevance labels per query-document pair~\citep{macavaney2023one, thomas2023large}.
Let $\hat{P}(R = r \mid d, q)$ indicate the predicted probability for relevance value $r$ for the combination of document $d$ and query $q$, the mean predicted relevance is then:
\begin{equation}
    \hat{\mu}(d) = \sum_{r \in \mathcal{R}}\hat{P}(R = r \mid d, q)\, r.
    \label{eq:predrelevance}
\end{equation}
Using these predicted relevances, we can construct a prediction of performance on the dataset-level from a sampled set of queries $Q$.
This results in the following predicted metric value:
\begin{equation}
    \hat{U}(Q) = \frac{1}{|Q|}\sum_{q \in Q}\sum_{d \in \mathcal{D}_q}\omega(\text{rank}(d \mid q,\mathcal{D}_q))\,\hat{\mu}(d \mid q).
    \label{eq:predictedutility}
\end{equation}
As discussed in previous work~\citep{faggioli2023perspectives,clarke20234}, basing $\hat{U}(Q)$ on state-of-the-art \acp{LLM} could greatly reduce costs~\citep{thomas2023large}, but there are many risks involved in replacing human annotators~\citep{bender2021dangers}.
The accuracy of $\hat{U}(Q)$ completely depends on the predictive capabilities of the generative model.
Thus, without further knowledge about the reliability of the predictions, one has no indication of its trustworthiness.
Our proposed methodologies use the available $n$ ground truth query-level performances together with the many generated relevance predictions to construct reliable \acp{CI} that quantify these risks.
\section{Method 1: Prediction-Powered Inference for Information Retrieval Evaluation}
\label{sec:method:ppi}

Our first proposed method applies the \acfi{PPI} framework to IR evaluation.
\ac{PPI} is a very recent advancement in \ac{CI} construction introduced by \citet{angelopoulos2023prediction}.
It utilizes computer-generated predictions to create smaller \ac{CI} when these predictions are somewhat accurate.
The core idea is to avoid estimating a variable on labelled data directly, and instead, build an estimate around the predictions which is then corrected based on the labelled data~\citep{breidt2017model}.
If the predictions are found to be accurate on the labelled data, then this increases our confidence that its predictions on unlabelled data are also accurate.
As predictions become available in much larger quantities, this can increase our confidence further in the overall estimate.
To the best of our knowledge, we are the first to apply \ac{PPI} to \ac{IR} evaluation.

\subsection{Classical empirical mean estimation}

Before we detail our application of \ac{PPI}, it is easiest to start with classical empirical estimates.
As stated in Section~\ref{sec:problemsetting}, our aim is to place a reliable \ac{CI} around the true performance $U(Q)$, and relevance annotations are available for the first $n$ queries in $Q$.
Therefore, we can make an empirical estimate of the mean metric performance based on these queries:
\begin{equation}
    \hat{U}^\text{emp}(Q) = \frac{\sum_{i=1}^n U(q_i)}{n},
    \;
    \hat{\sigma}_\text{emp}^2 = \frac{\sum_{i=1}^n\big(\hat{U}^\text{emp}(Q) - U(q_i)\big)^2}{n-1},
\end{equation}
where $\hat{\sigma}_\text{emp}^2$ is the estimated variance of the empirical estimate, and $U(q_i)$ the metric value for the single query $q_i$ (Eq.~\ref{eq:utility}).
We note that its variance is solely reflective of the ground truth data.
Obviously, this estimate does not fully utilize our problems setting, as it ignores the queries without ground truth relevance annotations and their corresponding computer-generated relevance annotations.

\subsection{Prediction-powered inference}

In contrast, \ac{PPI} mean estimation combines ground truth and predicted values to create an estimator that has potentially much lower variance.
In our setting, the \ac{PPI} estimator is a combination of the estimated mean predicted query performance and the estimated mean prediction error:
\begin{equation}
    \hat{U}^\text{PPI}(Q) =
    \underbrace{\frac{\sum_{i=1}^N\hat{U}(q_i)}{N}}_\text{mean prediction}
    +
    \underbrace{\frac{\sum_{i=1}^n U(q_i) - \hat{U}(q_i)}{n}}_\text{mean prediction error}.
\end{equation}
In other words, \ac{PPI} constructs an estimate of the query performance based on the predicted relevance annotations, and corrects it by the estimated error based on the difference between the predicted and ground truth annotations.
As a result, it is unbiased:
\begin{equation}
\mathbbm{E}_{Q \sim \mathcal{Q}}\big[\hat{U}^\text{PPI}(Q)\big] =
\mathbbm{E}_{Q \sim \mathcal{Q}}\big[\hat{U}^\text{emp}(Q)\big] =
 U(\mathcal{Q}),
\end{equation}
Whilst the empirical and \ac{PPI} estimates have the same expected value, the key-difference is their variances.
Assuming the queries are i.i.d., the estimated variance of \ac{PPI} can be decomposed into a part stemming from the mean prediction and another from the prediction error: 
\begin{align}
    \hat{\sigma}^2_\text{PPI}(Q) &= \hat{\sigma}^2_\text{pred}(Q) + \hat{\sigma}^2_\text{error}(Q), \nonumber
    \\
    \hat{\sigma}^2_\text{pred}(Q) &= \sum_{i=1}^N \frac{( \hat{U}(q_i) - \frac{1}{N}\sum_{j=1}^N\hat{U}(q_j) )^2}{N-1},
    \\
    \hat{\sigma}^2_\text{error}(Q) &= \sum_{i=1}^n \frac{(U(q_i) - \hat{U}(q_i) - \frac{1}{n}\sum_{j=1}^n (U(q_j) - \hat{U}(q_j)))^2}{n-1}.
    \nonumber
\end{align}
This reveals how \ac{PPI} can benefit from predictions and unlabelled data.
We see that $\hat{\sigma}^2_\text{error}$ shrinks as predicted performances become more accurate, whilst $\hat{\sigma}^2_\text{pred}$ shrinks as more unlabelled data becomes available (as $N$ increases).
Comparing $\hat{\sigma}^2_\text{PPI}$ with $\hat{\sigma}^2_\text{emp}$ reveals that \ac{PPI} can give a lower variance estimate, but only if predictions are somewhat accurate.
Conversely, when they are inaccurate the variance could actually be greater.

Finally, in order to turn the estimated mean and variance into a \ac{CI}, we follow \citet{angelopoulos2023prediction} and assume $ \hat{U}_\text{PPI}(Q)$ follows a normal distribution.
The 95\% confidence interval is then:
\begin{equation}
\hat{U}_\text{high/low}(Q) = 
\hat{U}^\text{PPI}(Q) \pm 1.96\sqrt{
\frac{ \hat{\sigma}^2_\text{error} }{n}
+ \frac{ \hat{\sigma}^2_\text{pred} }{N}
}.
\label{eq:PPI}
\end{equation}
Accordingly, one can use a different z-score than 1.96 to choose a different level of confidence.
We note that this implicitly assumes the prediction error follows a symmetric distribution.

This concludes our description of our \ac{PPI} method.
Its biggest advantage is its simplicity and straightforward application, making it attractive for practical usage.
A limitation is that it only gives a \ac{CI} of the overall performance (dataset-level).
Therefore, \ac{PPI} cannot be used to place \ac{CI} around individual query performances, and similarly, it cannot vary its confidence for different queries.

\begin{figure*}[t]
    \centering
    \includegraphics{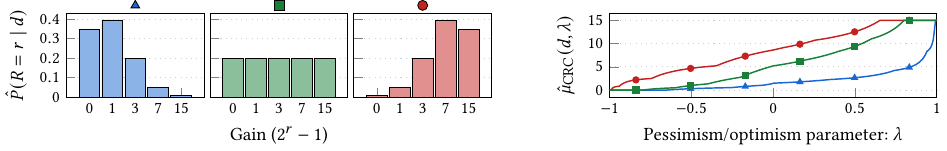}
    \vspace{-\baselineskip}
    \caption{
    Three different predicted relevance distributions (left) and their corresponding $\hat{\mu}_\text{CRC}(d, \lambda)$ curves (right).
    }
    \label{fig:CRCexample}
    \vspace{-\baselineskip}
\end{figure*}

\section{Background: Conformal Prediction and Conformal Risk Control}
\label{sec:background:conformal}

This section provides the necessary background on conformal prediction and \acf{CRC}~\citep{angelopoulos2021gentle,angelopoulos2022conformal, papadopoulos2008inductive, balasubramanian2014conformal}, before Section~\ref{sec:method:conformal} introduces our \ac{CRC} approach for IR evaluation.

\subsection{Conformal prediction}

Conformal prediction provides a unique approach to uncertainty quantification in predictions~\citep{papadopoulos2008inductive, balasubramanian2014conformal}.
The key characteristic of conformal prediction is that its predictions are not individual labels but \emph{sets} of labels.
For instance, the most basic version of this approach constructs a prediction set $\mathcal{C}$ by including all labels that have a predicted probability above a threshold $\lambda \in [0,1]$~\citep{angelopoulos2021gentle}.
Let $\hat{P}$ indicate a predicted probability, $X$ contextual features and $Y$ a corresponding label, this basic prediction set is then:
\begin{equation}
    \mathcal{C}_\text{basic}(X \mid \lambda) = \{y : \hat{P}(Y = y \mid X) > \lambda\}.
\end{equation}
Given a set of i.i.d.\ calibration data, conformal prediction can set $\lambda$ so that $\mathcal{C}_\text{basic}$ contains the true label with high probability:
\begin{equation}
    P\big(Y \in \mathcal{C}_\text{basic}(X \mid \lambda)\big)
    > 1 - \alpha.
    \label{eq:conformalprediction}
\end{equation}
Thereby, $\mathcal{C}_\text{basic}$ can capture the uncertainty in the prediction of $X$, with strong theoretical guarantees, when applied to the same distribution from which the calibration data was sampled~\citep{angelopoulos2021gentle}.

\subsection{Conformal risk control}

For purposes other than label prediction, there is a more general version of this approach: \acfi{CRC}~\citep{angelopoulos2022conformal}.
Let $\mathcal{C}(X \mid \lambda)$ be an arbitrary function that constructs sets that increase with $\lambda$, $\mathcal{L}$ a bounded loss function that shrinks as $\mathcal{C}$ grows, and in this context $\alpha \in \mathbb{R}$, \ac{CRC} aims to guarantee the expected loss is bounded:
\begin{equation}
    \mathbb{E}_{(x,y)\sim P(X,Y)}[\mathcal{L}(\mathcal{C}(X =x \mid \lambda), Y =y)] < 1 - \alpha.
    \label{eq:conformalriskguarantee}
\end{equation}
We can see that this is a generalized version of conformal prediction, since it is equivalent to Eq.~\ref{eq:conformalprediction} when:
$
    \mathcal{L}(\mathcal{C}, Y) = \mathbbm{1}[Y \not\in \mathcal{C}]
$~\citep{angelopoulos2021gentle}.

\ac{CRC} can guarantee Eq.~\ref{eq:conformalriskguarantee}, by finding a value of $\lambda$ based on a set of $n$ i.i.d.\ calibration data-points such that:
\begin{equation}
    \frac{1}{n}\sum_{i=1}^n \mathcal{L}(\mathcal{C}(X_i \mid \lambda), Y_i) < \alpha - \frac{B - \alpha}{n},
    \label{eq:conformalriskrequirement}
\end{equation}
where $B$ is the maximum possible value of $\mathcal{L}$.
Under the assumption that the calibration data was sampled from the same distribution ($P(X,Y)$), \ac{CRC} is proven to provide the bound guarantee stated in Eq.~\ref{eq:conformalriskguarantee}~\citep{angelopoulos2022conformal, angelopoulos2021gentle}.
We note that it is possible that no $\lambda$ value exists that can satisfy Eq.~\ref{eq:conformalriskrequirement} because the number of data-points $n$ is too small.
In this case, the method explicitly fails to provide a \ac{CI}, thereby, \ac{CRC} indicates when it is unable to guarantee reliable \acp{CI}.
The generality and flexibility of the \ac{CRC} framework enables us to build our own \ac{CI} method for IR evaluation on top of it.

\section{Method 2: Conformal Risk Control for Information Retrieval Evaluation}
\label{sec:method:conformal}
Our second proposed method uses \acfi{CRC} for \ac{CI} construction~\citep{angelopoulos2022conformal}.
In contrast with \ac{PPI}, it can provide both \ac{CI} around mean performance  and per query performance.
It also relies on different assumptions than \ac{PPI} and empirical bootstrapping.

Our description of the method is divided into three parts: firstly, we introduce our $\mathcal{C}$ function, secondly, we describe how calibration data is gathered, and thirdly, we propose our alternative dual-calibration approach specific for \acp{CI}.

\subsection{Optimistic and pessimistic estimation}
For our purposes, $\mathcal{C}$ will construct \acp{CI} for the relevance of each individual document, that are then translated into \acp{CI} on query and dataset-level performance.
Thus, our \ac{CRC} method treats each \ac{CI} as  a set that includes all values between its minimum and maximum.
Accordingly, we must predict the boundaries of \acp{CI} on a document-level,
therefore, we propose two functions: $\hat{\mu}_\text{high}$ and $\hat{\mu}_\text{low}$, that provide more optimistic and pessimistic predictions than $\hat{\mu}$, respectively.
We wish the optimism/pessimism to follow the confidence of the generative model,
thus,
we take the predicted distribution $\hat{P}$ and remove $\lambda$ probability from the top or bottom labels:\footnote{For brevity, we omit $q$ from our notation: $\hat{P}(R = r | d, q) = \hat{P}(R = r | d)$.}
\begin{equation}
\begin{split}
    \hat{Q}_\text{high}(R = r | d, \lambda)
    &=
    \hat{P}(R = r | d) - \max\!\Big(0, \lambda - \!\!\sum_{r' \in \mathcal{R} : r' < r \hspace{-0.75cm}} \hat{P}(R = r'| d)\!\Big),\!\!
    \\
    \hat{Q}_\text{low}(R = r | d, \lambda)
    &=
    \hat{P}(R = r | d) - \max\!\Big(0, \lambda - \!\!\sum_{r' \in \mathcal{R} : r' > r \hspace{-0.75cm}} \hat{P}(R = r'| d)\!\Big).\!\!
\end{split}
\end{equation}
We note that when $\lambda$ is greater than the predicted probability for the lowest/highest label, the remainder is subtracted from the next lowest/highest label, and so forth.
The results are normalized to produce the valid probability distributions; $\hat{P}_\text{high}$ and $\hat{P}_\text{low}$:
\begin{equation}
    \hat{P}_\text{high/low}(R = r \mid d, \lambda)
    =
    \frac{
    \hat{Q}_\text{high/low}(R = r \mid d, \lambda)
    }{
    \sum_{r'\in \mathcal{R}}\hat{Q}_\text{high/low}(R = r' \mid d, \lambda)
    }.
\end{equation}
Due to possible bias in the predicted relevance annotations, e.g., all predictions could be severe over or underestimates,
we want to enable both boundaries of \acp{CI} to be optimistic or pessimistic.
For elegance, we let $\lambda\in(-1,1)$ and our perturbed distribution is either optimistic or pessimistic based on the sign of $\lambda$:
\begin{equation}
    \hat{P}_\text{CRC}(R = r \mid d, \lambda) = 
    \begin{cases}
    \hat{P}_\text{high}(R = r \mid d, \lambda) & \text{if } \lambda \geq 0,
    \\
    \hat{P}_\text{low}(R = r \mid d, -\lambda) & \text{otherwise.}
    \end{cases}
\end{equation}
The final optimistic or pessimistic estimates are the expected values over these perturbed distributions:
\begin{equation}
    \hat{\mu}_\text{CRC}(d ,\lambda) = \sum_{r \in \mathcal{R}}\hat{P}_\text{CRC}(R = r \mid d, \lambda)r.
\end{equation}

Figure~\ref{fig:CRCexample} visualizes how $\hat{\mu}_\text{CRC}$ varies over different $\lambda$ values for three different predicted relevance distributions.
We see that low predicted probabilities for the largest labels mean that $\lambda$ has to be greater for $\hat{\mu}_\text{CRC}$ to reach high values, and vice versa, $\lambda$ has to be lower for low probabilities for the lowest label values to reach low values.
In other words, it takes more extreme $\lambda$ values for $\hat{\mu}_\text{CRC}$ to be heavily optimistic when the generative model is very confidently pessimistic, and vice versa.

The document-level $\hat{\mu}_\text{CRC}$ are translated to performance estimates following Eq.~\ref{eq:utility}~\&~\ref{eq:predictedutility} but with $\hat{\mu}(d)$ replaced by $\hat{\mu}_\text{CRC}(d,\lambda)$.
Finally, to construct \acp{CI}, we use two parameters: $\lambda_\text{high} \in (-1,1)$ and $\lambda_\text{low} \in (-1,1)$, s.t.\ $\lambda_\text{low}<\lambda_\text{high}$, to obtain $\hat{U}_\text{CRC}(Q, \lambda_\text{low})$ and $\hat{U}_\text{CRC}(Q, \lambda_\text{high})$.
The predicted \ac{CI} is the range between the perturbed estimates:
\begin{equation}
    \mathcal{C}(Q, \lambda_\text{high}, \lambda_\text{low}) = [\hat{U}_\text{CRC}(Q, \lambda_\text{low}), \hat{U}_\text{CRC}(Q, \lambda_\text{high})].
    \label{eq:CCRC}
\end{equation}
Our proposed $\mathcal{C}$ function has several significant properties that enable it to function well as \ac{CI}:
When $\lambda_\text{low}=\lambda_\text{high}=0$, it only contains the predicted $\hat{U}(Q)$ value, since:
%
$
    \hat{U}_\text{CRC}(Q,0) = \hat{U}(Q)
$.
%
As the $\lambda$ approach one and minus one, the perturbed estimates become the minimum and maximal possible metric values:
\begin{equation}
   \lim_{\lambda_\text{high}\to1, \lambda_\text{low}\to-1}
   \mathcal{C}(Q, \lambda_\text{high}, \lambda_\text{low}) =
   [\max U(\cdot),  \min U(\cdot)].
\end{equation}
Consequently, there always exists values for $\lambda_\text{high}$ and $\lambda_\text{low}$ to bound the true performance $U(\mathcal{Q})$, since it must lie between the minimal and maximal possible metric values:
\begin{equation*}
 \exists \lambda_\text{high} \in (-1,1), \lambda_\text{low} \in (-1,\lambda_\text{high}];\quad
 U(\mathcal{Q}) \in \mathcal{C}(Q, \lambda_\text{high}, \lambda_\text{low}).
\end{equation*}

To summarize, we have proposed a novel $\mathcal{C}(Q, \lambda)$ function that creates a \ac{CI} based on the relevance annotations of a generative model.
It follows the confidence of the underlying generative model by perturbing the predicted relevance distributions in an optimistic or pessimistic manner.
The remainder of this section explains how we determine the values of $\lambda_\text{high}$ and $\lambda_\text{low}$ such that a reliable \ac{CI} is found that captures the true metric value with high confidence.

\subsection{Data sampling and bootstrapping}
\label{sec:datasampling}
In order to perform \ac{CRC} calibration, a set of ground truth examples is required to serve as calibration data.
In our setting, we aim to estimate the mean over the true query-distribution $\mathcal{Q}$ based on the sampled set of queries $Q$.
Accordingly, a set of examples of mean estimates based on sampled set from $\mathcal{Q}$ is required;
we create $M$ examples by sampling from the $n$ queries in $Q$ with ground truth annotations: $\bar{Q}_i \subset \{q_1,q_2,\ldots q_n\}$.
The collection of these $M$ sets should mimic the distribution of $\mathcal{Q}$:
%
$
\bar{\mathcal{Q}} = \{\bar{Q}_1, \bar{Q}_2, \ldots, \bar{Q}_M \}.
$
%

There are many options to construct $\bar{\mathcal{Q}}$, for instance, one could sample queries with or without replacement, the size of the sampled sets could be varied, etc.
Moreover, if one wants to create \acp{CI} around the performance of each query, they can choose the sets to contain a single query: $\bar{Q}_i = \{q_i\}$.
Another option is to sample queries and subsets of the document to be ranked, to artificially increase the variety in candidate documents available per query.
Choices that increase the number of examples $M$ have the potential to decrease \ac{CI} width.
However, if the resulting $\bar{\mathcal{Q}}$ is no longer representative of the true distribution $\mathcal{Q}$, the reliability of the \acp{CI} will decrease.

\subsection{Dual-calibration for confidence intervals}

With our definition of $\mathcal{C}(Q, \lambda_\text{high}, \lambda_\text{low})$ and the calibration data $\bar{\mathcal{Q}}$, all that remains is to calibrate $\lambda_\text{high}$ and $\lambda_\text{low}$.
However, standard \ac{CRC} is designed for the calibration of a single parameter.
Luckily, for the purpose of construction a \ac{CI}, we can apply \ac{CRC} calibration sequentially.
Because for any $\hat{U}_\text{low} < \hat{U}_\text{high}$, the following holds:
\begin{equation}
\begin{split}
&\Big(
P(U \leq \hat{U}_\text{low}) \leq \frac{\alpha}{2}
\land
P(U \geq \hat{U}_\text{high}) \leq \frac{\alpha}{2}
\Big)
\\&\qquad\qquad\qquad\qquad
\longrightarrow P(\hat{U}_\text{low} \leq U \leq \hat{U}_\text{high})
\leq 1 - \alpha.
\end{split}
\label{eq:dualrequirement}
\end{equation}
Therefore, we can first calibrate one of the bounds with \ac{CRC}, and the other afterwards.
Accordingly, we propose two loss functions:
\begin{equation}
\begin{split} 
\mathcal{L}_\text{high}\big(\mathcal{C}_\text{CRC}(Q, \lambda_\text{high}), U(Q)\big)
&= 
 \mathbbm{1} \big[ \hat{U}_\text{CRC}(Q, \lambda_\text{high}) < U(Q) \big],\!
\\
\mathcal{L}_\text{low}\big(\mathcal{C}_\text{CRC}(Q, \lambda_\text{low}), U(Q)\big)
&= \mathbbm{1} \big[
 \hat{U}_\text{CRC}(Q, \lambda_\text{low}) > U(Q) \big].
 \end{split}
\end{equation}
Through applying two binary search procedures, we find the values for $\lambda_\text{high} \in (-1,1)$ and $\lambda_\text{low} \in (-1,1)$ such that $\lambda_\text{low} < \lambda_\text{high}$ and:
\begin{equation}
\frac{1}{M} \! \sum_{i=1}^M \! \mathcal{L}_\text{high/low}\big(\mathcal{C}_\text{CRC}(\bar{Q}_i, \lambda_\text{high/low}), U(\bar{Q}_i)\big) \! < \! \frac{1}{2}\mleft(\!\alpha - \frac{1 \!-\! \alpha}{M}\!\mright).
\label{eq:dualcalibration}
\end{equation}
Consequently, according to Eq.~\ref{eq:dualrequirement}, it must be the case that the \ac{CRC} requirement for the complete interval holds:
\begin{equation}
\frac{1}{M} \sum_{i=1}^M 
\mathbbm{1}\mleft[
U(\bar{Q}_i) \in \mathcal{C}_\text{CRC}(\bar{Q}_i, \lambda_\text{high}, \lambda_\text{low})
\mright]
< \alpha - \frac{1 - \alpha}{M}.
    \label{eq:dualcrcrequirement}
\end{equation}
Therefore, the resulting \ac{CI} has the desired reliability, when applied to the distribution underlying $\mathcal{\bar{Q}}$:
\begin{equation}
P\Big(
    \hat{U}_\text{CRC}(\bar{Q}, \lambda_\text{low})
    \!\leq\!
    U(\bar{Q})
    \!\leq\! 
    \hat{U}_\text{CRC}(\bar{Q}, \lambda_\text{high})
    \,|\, \bar{Q} \sim \bar{\mathcal{Q}}
    \Big) 
    > 1 - \alpha.
\label{eq:CRCsampledrequirement}
\end{equation}
Accordingly, we assume that this (Eq.~\ref{eq:CRCsampledrequirement}) implies the following:
\begin{equation}
P\mleft(
\hat{U}_\text{CRC}(Q, \lambda_\text{low}) \!\leq\! U(\mathcal{Q}) \!\leq\! \hat{U}_\text{CRC}(Q, \lambda_\text{high}) \,|\, Q \sim \mathcal{Q} \mright)
> 1 - \alpha.
\label{eq:crcguarantee}
\end{equation}
This is a very standard assumption made in \ac{CI} literature, and at the core of many bootstrapping methods~\citep{diciccio1996bootstrap, efron1987better}.
If $\bar{\mathcal{Q}}$ is created by standard sampling from $Q$, then this is a relatively safe assumption.

\subsection{Overview}

Finally, we give an overview of the different components in our \ac{CRC} approach:
Our \ac{CI} are created with the $\mathcal{C}_\textrm{CRC}(Q, \lambda_\text{high}, \lambda_\text{low})$ function (Eq.~\ref{eq:CCRC}), where $Q$ are all available queries (no ground truth annotations required).
We note that when the set $Q$ contains a single query, it produces a \ac{CI} for query-level performance.

The resulting \ac{CI} are only reliable if $\lambda_\text{high}$ and $\lambda_\text{low}$ are properly calibrated.
We do so by first sampling a collection of query-sets $\bar{\mathcal{Q}}$ (Section~\ref{sec:datasampling})
and calibrating each parameter independently (Eq.~\ref{eq:dualcalibration}).
Due to the nature of \ac{CI} (Eq.~\ref{eq:dualrequirement}), this guarantees the \ac{CRC} requirement is met (Eq.~\ref{eq:dualcrcrequirement}), and assuming $\bar{\mathcal{Q}}$ is representative of $\mathcal{Q}$, this guarantees that our \ac{CI} are reliable with a given probability (Eq.~\ref{eq:crcguarantee}).

\begin{table}
    \centering
    \caption{DCG@10 performance of different rankers as measured by human-annotated labels and LLM-generated labels. Each approach ranks the top-100 results retrieved by BM25.}
    \vspace{-0.75\baselineskip}
    \begin{tabular}{lrrrr}
    \toprule
     & \multicolumn{2}{c}{TREC-DL} & \multicolumn{2}{c}{Robust04}  \\
     & Human & LLM & Human & LLM \\
    \midrule
    Random  &  3.16 & 6.86  & 0.99 & 2.96 \\
    BM25    &  8.25 & 12.93 & 2.71 & 4.32 \\
    LLM     & 12.81 & 23.73 & 3.23 & 7.17 \\
    Perfect & 19.00 & 17.44 & 5.70 & 4.65 \\
    \bottomrule
    \end{tabular}
    \label{tab:datasets}
\end{table}

\section{Experimental Setup}
Our experiments compare the confidence intervals produced by \ac{PPI}, \ac{CRC} and classical empirical bootstrapping on benchmark \ac{IR} datasets, 
by answering the following research questions:%
\footnote{
Our experimental implementation and our dataset of generated LLM labels are available at:
\url{https://github.com/google-research/google-research/tree/master/high_confidence_ir_eval_using_genai}
}
\begin{itemize}
\item[\textbf{RQ1}:]How many human-annotated labels are required to produce informative confidence intervals?
\item[\textbf{RQ2}:]How resilient are the confidence intervals to systematic mistakes made by LLM labelers?
\item[\textbf{RQ3}:]What benefits could PPI and CRC get from potential improvements in the accuracy of label generation?
\item[\textbf{RQ4}:]Can CRC capture differences in uncertainty per query?
\end{itemize}

\textbf{LLM-generated relevance labels.}
For each query-document pair, a prompt is constructed that asks the LLM to assess the relevance according to the relevance scales of the dataset, in our case: 0--2 (Robust04) and 0--3 (TREC-DL).
The LLM is provided with clear definitions of the different relevance labels, similar to~\cite{thomas2023large}.
Specifically, instructions that give definitions for relevance labels in each prompt.
We chose prompts that mimic the instructions for human annotators as closely as possible, hereby, we hope to precisely simulate the manual labeling process for each dataset.
The exact prompts are provided in Appendix~\ref{appendix:prompts}.

To obtain relevance labels, we run the LLMs in `\emph{scoring mode}'~\cite{zhuang2023beyond}.
That is, for each relevance label $r \in \mathcal{R}$, we compute the log-probability of the LLM outputting the relevance rating $r$.
The log-probabilities are then normalized via a softmax function so that we obtain a probability distribution that represents the LLM's confidence in assigning each relevance label to the query-document pair.

As our LLM model, we choose to use Flan-UL2~\cite{tay2022unifying,longpre2023flan}, because it is open source and has demonstrated strong performance on ranking tasks~\cite{qin2023large}.
It is worth noting that larger, more powerful, LLMs exist~\citep{team2023gemini}, and that we do not utilize any prompt-engineering~\citep{thomas2023large}.
These choices were made because the goal of our experiments is not to find the best LLM-generated labels, but to confirm whether the confidence intervals proposed by our methods accurately capture the uncertainty in LLM-generated relevance labels.
Since advancements in LLM techniques result in rapid changes in the state-of-the-art, we choose to focus on the established human annotator setting instead~\citep{craswell2021trec, voorhees2005trec}.

\begin{figure}
    \centering
    \includegraphics{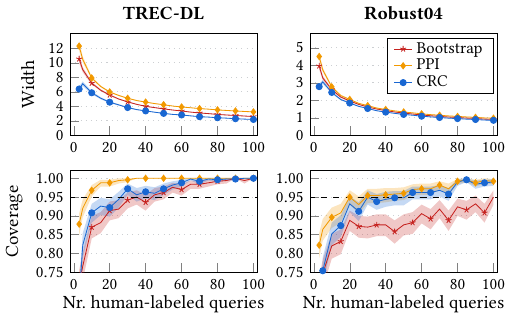}
    \vspace{-2\baselineskip}
    \caption{Width (top) and coverage (bottom) of the confidence intervals produced by the methods. The dashed line in the bottom plots is the 95\% coverage target. Shaded areas indicate 95\% prediction intervals over 500 independent runs.}
    \label{fig:widthandcoverage}
    \vspace{-1.25\baselineskip}
\end{figure}

\textbf{Datasets.}
Our evaluation is based on two established benchmark datasets: TREC-DL~\citep{craswell2021trec} and TREC-Robust04~\citep{voorhees2005trec}.
Both datasets are comprised of documents and queries together with human-annotated relevance judgments.
For each dataset, we perform a random 50:50 split to obtain a validation and test set where the validation set is used for calibration of the methods.
(A training set is not required in our setting.)
To avoid distribution shifts, for TREC-DL, we create a stratified sample over four years (2019 - 2022) that ensures each year is equally represented in each split.
As the ranker to evaluate, we choose BM25, as the metric we choose DCG@10~\citep{jarvelin2002cumulated}.
In other words, our methods will construct \acp{CI} around the DCG@10 of BM25 on both datasets.
Table~\ref{tab:datasets} displays the ranking performance of BM25 and the LLM-generated labels.
To match the gain function of DCG all labels were transformed accordingly: $r' = 2^{r}-1$, for all performance estimations.

\textbf{Methods in comparison.}
The methods included in our comparison are:
\begin{enumerate*}[label=(\roman*)]
\item empirical bootstrapping~\citep{diciccio1996bootstrap},
\item \acfi{PPI} (Section~\ref{sec:method:ppi}), and
\item \acfi{CRC} (Section~\ref{sec:method:conformal}).
\end{enumerate*}
The empirical bootstrap approach acts as a baseline that only considers the available human-labeled data, this is a standard approach in previous \ac{IR} literature~\citep{webber2010measurement, webber2013approximate, aslam2006statistical, yilmaz2008simple,sakai2014statistical}.
All our empirical bootstrap \ac{CI} are based on 10,000 bootstrap samples.
\ac{PPI} is computed by applying Eq.~\ref{eq:PPI} to both the validation set (the first $n$ queries) and the test set (the remaining $N-n$ queries), it utilizes both human and LLM-generated labels.
Finally, our CRC approach also utilizes both, we use the validation set to calibrate the $\lambda$ parameters and then compute the \ac{CI} using only the LLM-generated labels on the test-set.
For calibration, CRC is provided $M=$~10,000 batches each consisting of $n$ queries that were sampled with replacement from the validation set (see Section~\ref{sec:datasampling}).
We note that the batch size depends on the number of available queries with human annotations, which is varied in our experiments.
For the \acp{CI} to be evaluated, the \ac{CI} is applied to the entire test-set to obtain a dataset-level \ac{CI}, i.e., we compute $\mathcal{C}(Q_\text{test}, \lambda_\text{high}, \lambda_\text{low})$ (Eq.~\ref{eq:CCRC}).
Some of our experiments consider CRC \acp{CI} around query-level performance, in these cases, $\lambda$ is not calibrated on bootstrapped batches but on $n$ batches that each contain a single query.

We evaluate the \acp{CI} produced by each method by considering their \emph{width} and \emph{coverage}.
The width measures how wide and thus how informative or specific the \ac{CI} is, where a smaller width is better.
The coverage measures how frequent the \ac{CI} covers the true performance on the test-set over 500 independently repeated experiment runs, thus the higher the better.
The target for all the methods is a coverage of 95\% or higher and we set $\alpha = 0.05$ accordingly.

\begin{figure}
    \centering
    \includegraphics{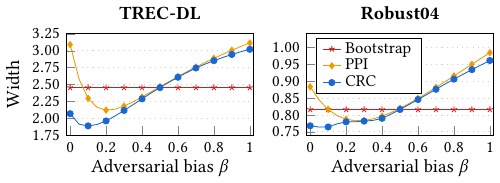}
    \includegraphics{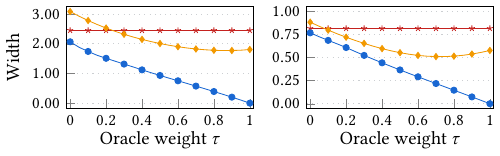}
    \vspace{-2\baselineskip}
    \caption{Width of the confidence intervals for increasing levels of LLM bias ($\beta$, top-row) and oracle-enhanced LLM accuracy ($\tau \rightarrow 1$, bottom row) with $n=112$ on TREC-DL and $n=125$ on Robust04.  Shaded areas indicate 95\% prediction intervals over 500 independent runs.
    Coverage plots are omitted since all methods maintain >95\% coverage.
    }
    \label{fig:noise}
    \label{fig:oracle2}
    \vspace{-\baselineskip}
\end{figure}

\begin{figure*}
    \centering
    \includegraphics{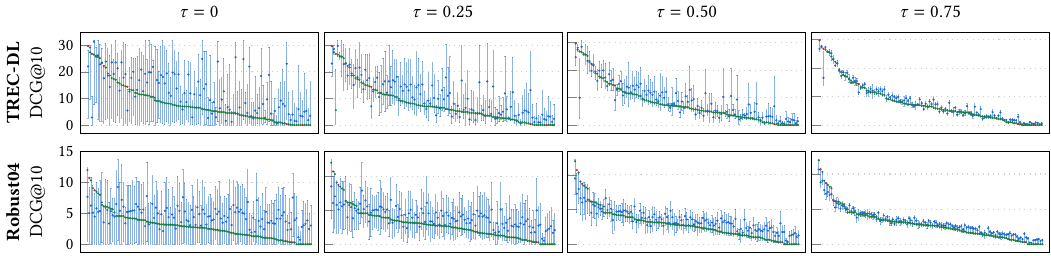}
    \vspace{-\baselineskip}
    \caption{
    95\% \ac{CI} produced per-query by CRC using LLM predicted relevance annotations ($\tau = 0$) and oracle-enhanced LLM annotations ($\tau > 0$).
    The queries are sorted by their true DCG performance (according to human-annotations), indicated by red and green dots.
    Green dots are covered by their \ac{CI} whereas red dots are not.
    Blue dots indicate the predicted DCG performance (according to LLM-generated annotations).
    Clearly, the \acp{CI} shrink considerably as annotations become more accurate ($\tau \rightarrow 1$).
    }
    \label{fig:confintervals}
    \label{fig:oracle}
    \vspace{-\baselineskip}
\end{figure*}

\section{Results}

\subsection{Number of required human-annotations}

Our main results are displayed in Figure~\ref{fig:widthandcoverage}.
Here we see how the width and coverage of the different methods vary, as they are provided with $n$ queries with human annotations sampled from the validation set.
As expected, all methods provides better \acp{CI} when provided with larger portion human-annotated queries, i.e., as $n$ increases coverage increases and width decreases.

We start by considering the performance of the empirical bootstrap baseline.
On the TREC-DL dataset, we see that it requires at least 40 labeled queries to achieve 95\% coverage.
Furthermore, on Robust04, with 100 labeled queries it almost reaches 95\% coverage. However, the plotted prediction intervals around the reported coverage reveal that many of its runs did not reach 95\% coverage.

In contrast, both our PPI and CRC approaches have stronger coverage with less queries:
PPI needs less than 20 queries on TREC-DL and less than 40 on Robust04.
Similarly, CRC needs less than 30 on TREC-DL and less than 50 on Robust04.
In terms of width, CRC clearly provides the smallest width of all the methods, whilst PPI is worse than empirical bootstrap on TREC-DL and comparable on Robust04.
This comparison is not entirely fair, i.e., there is generally a tradeoff between coverage and width, it appears PPI does better in terms of coverage but that results in wider \acp{CI}.
Thus, PPI has a clear advantage over empirical bootstrap on Robust04 where it has the same width but much better coverage.
Nevertheless, when CRC and PPI have the same coverage, CRC has smaller widths, with an especially large difference on TREC-DL.
Therefore, it appears that CRC has the most informative \ac{CI}, whilst PPI needs fewer queries to reach 95\% coverage.
Both methods provide substantial improvements over empirical bootstrapping.

Thus we answer RQ1 as follows: both PPI and CRC require as few as 30 human-labeled queries to produce informative and reliable confidence intervals.
Whilst empirical bootstrapping requires significantly more human-labeled queries to achieve similar results.

\subsection{Sensitivity to LLM accuracy}
Our PPI and CRC methods can benefit from accurate LLM labels, but in order to be reliable, it is also important that they are robust to inaccurate labels.
We investigate the effect of LLM accuracy by adding adversarial bias to the predicted relevance distributions, with $\beta \in [0, 1]$, change the predictions as follows:
$
\hat{P}_{\beta}(R = r |\, d, q) = \frac{1}{Z} \mleft((1-\beta) \hat{P}(R = r |\, d, q) + \beta (1 - \hat{P}(R = r |\, d, q)) \mright),
$
where $Z$ is a normalizing factor to ensure the result is a valid probability distribution. For $\beta = 0$ this leaves predictions unaltered, with $\beta = 0.5$ this is a uniform distribution and at $\beta = 1$ it produces the inverse of the original predictions.

Figure~\ref{fig:noise} shows the widths as $\beta$ is varied ($n=112$ on TREC-DL and $n=125$ on Robust04).
We do not report coverage as all methods obtain a mean coverage of at least 95\%.
When $\beta<0.5$, CRC consistently provides better widths than empirical bootstrap, whilst PPI has inconsistent improvements.
As expected, when $\beta>0.5$ both methods do worse than empirical bootstrap in terms of width.

Thus, we can answer RQ2: the coverage of both PPI and CRC-bootstrap are robust to systematic mistakes made by the LLM, however, improvements in widths are dependent on LLM accuracy.

\subsection{Potential from more accurate labels}
\label{sec:results:oracle}
We run additional experiments to understand how the \acp{CI} behave under an oracle LLM: one that can perfectly generate relevance labels.
In Figure~\ref{fig:oracle2}, we increasingly interpolate between the LLM-generated relevance labels and the true (human-annotated) relevance labels using a parameter $\tau \in [0,1]$.
As $\tau$ increases, the performance of the LLM labels becomes better.
First, we note that all methods retain a perfect 100\% coverage in these scenarios, so we omit the plots for coverage.
The empirical bootstrap approach does not use the LLM-generated labels and its CI is thus not impacted by the increasingly stronger LLM labels.
The PPI method is able to leverage the stronger LLM labels and is able to significantly outperform the empirical bootstrap method.
The fact that its CI is placed around the overall performance (dataset-level), prevents it from further improving the width, as it is inherently limited by the number of queries.
The CRC approaches are able to work around this limitation by efficiently identifying that the LLM-generated labels are more accurate as $\tau \rightarrow 1$ on the \emph{per-document} level.
Their per-query CIs correspondingly shrink and approach 0 as the LLM-generated labels become better.
This answers RQ3: Both PPI and CRC benefit from improvements in label generation accuracy.

\subsection{Query-performance confidence intervals}

We plot the confidence intervals produced by CRC on individual queries in Figure~\ref{fig:confintervals}.
Each plot in the figure shows the true DCG (based on human-annotated relevance labels) and the predicted DCG (based on LLM-generated labels) of all queries in the test split.
The queries are sorted by their true DCG, that is, queries where the ranker performs best appear on the left and progressively the query performance goes down.
Furthermore, we plot the per-query CI for varying values of $\tau$, to indicate how the confidence intervals behave as the LLM-generated labels become more accurate, similar to Section~\ref{sec:results:oracle}.
First, for all plots, we observe that the CIs vary per query: CRC captures the  uncertainty throughout LLM-generated labels.
Second, for the LLM-generated labels ($\tau = 0$), we observe that when the LLM predicts that the ranker performs poorly on a query, the bounds tend to be smaller for that query.
Similarly, when the predicted performance of the ranker is large, the bounds tend to be wider.
This indicates that the LLM-generated labels are generally better at identifying queries with poor ranking performance.
Third, as $\tau \rightarrow 1$, we see that CRC is able to identify that the labels are more accurate and its per-query CIs become significantly tighter.
This shows that CRC is not only able to vary its CI per query, but is also able to establish better per-query CIs as LLM labels become more accurate.
This is especially noticeable in the $\tau = 0.75$ plot for TREC-DL (top-right plot in Figure~\ref{fig:oracle}).
In this plot there is a single outlier query on the left where the LLM is wrong and its predicted labels are uncertain.
As a result the CRC method correctly places a very wide CI around this particular query, while keeping the CIs on other queries tight.
Finally, on both datasets the empirical coverage of 95\% is reached, indicating the CIs are reliable.
Thus, we answer RQ4 positively: CRC is able to construct CIs on a \emph{per-query} basis.

\section{Conclusion}
In this paper we study reliable evaluation of \ac{IR} systems using \ac{LLM}-generated relevance labels.
Obtaining human-annotated relevance labels is costly, especially in low-resource settings.
While \acp{LLM} can help generate relevance labels at scale, they are prone to make systematic errors and may be unreliable.
We resolve this by introducing two methods that construct \acfp{CI} around ranking metrics produced by \ac{LLM}-generated relevance labels: PPI and CRC.
These approaches require a small amount of reliable ground truth annotations to statistically analyze the distribution of errors and correct those errors.

Our results demonstrate that the proposed methods can correct errors in \ac{LLM}-generated labels and produce reliable \acp{CI}. Compared to other \ac{CI} approaches, we can produce \acp{CI} of superior coverage with tighter bounds, leading to more informative evaluation. Furthermore, the \acp{CI} produced by CRC can be computed per-query, providing further insights into low or high performing queries.

Our work is not without limitations.
First, we note that our methods require an LLM with scoring-mode to produce a distribution over LLM labels.
For LLMs without scoring-mode one could generate multiple labels stochastically to approximate a predicted distribution.
Second, our results suggest that applying some smoothing to the LLM-generated label distribution is beneficial to the resulting \acp{CI}.
How to systematically optimize the amount of smoothing is an open question.
Similarly, fine-tuning or prompt-engineering could also lead to distributions better suited for \ac{CI} construction.
Third, we only use the Flan-UL2 as an LLM labeler.
Our work can be extended to use different and potentially more powerful LLMs.
Future work could explore all of these directions further.

\section*{Acknowledgements}
This research was supported by the Google Visiting Researcher program.
Any opinions, findings and recommendations expressed in this work are those of the authors and are not necessarily shared or endorsed by their respective employers or sponsors.

\bibliographystyle{ACM-Reference-Format}
\citestyle{acmauthoryear}
\setcitestyle{numbers,sort&compress}

\bibliography{references}
\balance

\appendix
\section{Prompts}
\label{appendix:prompts}

The exact prompts used in our experiments are listed here.
We note that these prompts are tailored towards each dataset and use the relevance label definitions that human labelers used for each dataset. The \texttt{\{query\}} and \texttt{\{passage\}}/\texttt{\{document\}} are placeholders that are formatted with the actual query and passage/document during inference.

We observed that the model is sensitive to the particular prompt and dataset during scoring mode. For TREC-DL we score the suffixes "\texttt{0}", "\texttt{1}", "\texttt{2}" and "\texttt{3}". For Robust04 we found that scoring the suffixes with brackets is more effective: "\texttt{[0]}", "\texttt{[1]}" and "\texttt{[2]}".


\lstset{
  basicstyle=\footnotesize\ttfamily,
  columns=fullflexible,
  frame=single,
  breaklines=true,
  escapeinside={<@}{@>}
}

\begin{minipage}{\linewidth}
\begin{lstlisting}[caption={Prompt for TREC-DL.}]
Assess the relevance of the passage to the query on a four-point scale:
[0] Irrelevant: The passage has nothing to do with the query.
[1] Related: The passage seems related to the query but does not answer it.
[2] Highly relevant: The passage has some answer for the query, but the answer may be a bit unclear, or hidden amongst extraneous information.
[3] Perfectly relevant: The passage is dedicated to the query and contains the exact answer.

Query: <@\textcolor{red}{\{query\}}@>
Passage: <@\textcolor{red}{\{passage\}}@>
Relevance:
\end{lstlisting}
\end{minipage}

\begin{minipage}{\linewidth}
\begin{lstlisting}[caption={Prompt for Robust04.}]
Assess the relevance of the document to the query on a three-point scale:
[0] Not relevant: The document is not relevant to the query.
[1] Relevant: Parts of the document may be relevant to the query.
[2] Highly Relevant: The document is highly relevant to the query.

Query: <@\textcolor{red}{\{query\}}@>
Document: <@\textcolor{red}{\{document\}}@>
Relevance:
\end{lstlisting}
\end{minipage}

\end{document}